\documentclass[]{aa}  
\usepackage{ae,aecompl}
\usepackage{adjustbox}
\usepackage{graphicx}
\usepackage{xspace}
\usepackage{txfonts}
\usepackage{hyperref}
\usepackage{xcolor}
\usepackage{float}

\hypersetup{colorlinks,linkcolor={blue},citecolor={blue},urlcolor={blue}}  
\usepackage{diagbox}
\usepackage{epsf,palatino,url,wrapfig,array,setspace,amsmath,amssymb,fancyhdr,multirow,lscape,appendix,rotating}
\usepackage{graphics,epsfig,txfonts}
\usepackage{graphicx}
\usepackage{threeparttable}
 
\newcommand{\hsp}{3HSP~J095507.9+355101}

\newcommand{\fermi}{\emph{Fermi}}

\newcommand{\lehamoc}{{\tt LeHaMoC}\xspace}
\newcommand{\atheva}{{\tt ATHE$\nu$A}\xspace}
\newcommand{\nn}{NN\xspace}
\newcommand{\nns}{NNs\xspace}
\newcommand{\ml}{ML\xspace}

\usepackage{xparse,xcoffins}

\ExplSyntaxOn
\NewCoffin\imagecoffin
\NewCoffin\labelcoffin

\keys_define:nn { miguel/label }
 {
  label   .tl_set:N = \l_miguel_label_tl,
  labelbox .bool_set:N = \l_miguel_label_box_bool,
  labelbox .default:n = true,
  fontsize .tl_set:N = \l_miguel_label_size_tl,
  fontsize .initial:n = \footnotesize,
  pos .choice:,
  pos/nw .code:n = \tl_set:Nn \l_miguel_label_pos_tl { left,up },
  pos/ne .code:n = \tl_set:Nn \l_miguel_label_pos_tl { right,up },
  pos/sw .code:n = \tl_set:Nn \l_miguel_label_pos_tl { left,down },
  pos/se .code:n = \tl_set:Nn \l_miguel_label_pos_tl { right,down },
  pos/n .code:n = \tl_set:Nn \l_miguel_label_pos_tl { hc,up },
  pos/w .code:n = \tl_set:Nn \l_miguel_label_pos_tl { left,vc },
  pos/s .code:n = \tl_set:Nn \l_miguel_label_pos_tl { hc,down },
  pos/e .code:n = \tl_set:Nn \l_miguel_label_pos_tl { right,vc },
  pos .initial:n = nw,
  unknown .code:n   = \clist_put_right:Nx \l_miguel_label_clist
                       { \l_keys_key_tl = \exp_not:n { #1 } }
 }
\clist_new:N \l_miguel_label_clist
\box_new:N \l_miguel_label_box
\box_new:N \l_miguel_label_image_box

\NewDocumentCommand{\xincludegraphics}{O{}m}
 {
  \group_begin:
  \tl_clear:N \l_miguel_label_tl
  \clist_clear:N \l_miguel_label_clist
  \keys_set:nn { miguel/label } { #1 }
  \tl_if_empty:NTF \l_miguel_label_tl
   {
    \miguel_includegraphics:Vn \l_miguel_label_clist { #2 }
   }
   {
    \SetHorizontalCoffin\imagecoffin
     {
      \miguel_includegraphics:Vn \l_miguel_label_clist { #2 }
     }
    \SetHorizontalCoffin\labelcoffin
     {
      \raisebox{\depth}
       {
        \bool_if:NTF \l_miguel_label_box_bool
         { \fcolorbox{white}{white}{\l_miguel_label_size_tl\l_miguel_label_tl} }
         { \l_miguel_label_size_tl\l_miguel_label_tl }
       }
     }
    \SetVerticalPole\imagecoffin{left}{25pt+\CoffinWidth\labelcoffin/2}
    \SetVerticalPole\imagecoffin{right}{\Width-3pt-\CoffinWidth\labelcoffin/2}
    \SetHorizontalPole\imagecoffin{up}{\Height-5pt-\CoffinHeight\labelcoffin/2}
    \SetHorizontalPole\imagecoffin{down}{3pt+\CoffinHeight\labelcoffin/2}
    \use:x{\JoinCoffins\imagecoffin[\l_miguel_label_pos_tl]\labelcoffin[vc,hc]} 
    \TypesetCoffin\imagecoffin
   }
   \group_end:
 }
\NewDocumentCommand{\setlabel}{m}
 {
  \keys_set:nn { miguel/label } { #1 }
 }

\cs_new_protected:Nn \miguel_includegraphics:nn
 {
  \includegraphics[#1]{#2}
 }
\cs_generate_variant:Nn \miguel_includegraphics:nn { V }

\ExplSyntaxOff

\graphicspath{{./}{plots_arxiv/}}
\begin{document}

   \title{Application of neural networks to synchro-Compton blazar emission models} 
   \titlerunning{Neural networks for blazars}
   \author{A. Tzavellas\inst{1}\thanks{\email{tzavellas.anastasios@gmail.com}},
          G. Vasilopoulos\inst{1,2}\thanks{\email{gevas@phys.uoa.gr}},
          M. Petropoulou\inst{1,2},            
          A. Mastichiadis\inst{1},
          S.~I. Stathopoulos\inst{1,2}
          }       
   \institute{Department of Physics, National and Kapodistrian University of Athens, University Campus Zografos, GR 15784, Athens, Greece
 \and
    Institute of Accelerating Systems \& Applications, University Campus Zografos, Athens, Greece}
   \date{Received ... ; accepted ...}

 
  \abstract
   {Jets from supermassive black holes in the centers of active galaxies are the most powerful persistent sources of electromagnetic radiation in the Universe. To infer the physical conditions in the otherwise out-of-reach regions of extragalactic jets we usually rely on fitting of their spectral energy distribution (SED). The calculation of radiative models for the jet non-thermal emission usually relies on numerical solvers of coupled partial differential equations.} 
   {In this work machine learning is used to tackle the problem of high computational complexity in order to significantly reduce the SED model evaluation time, which is needed for SED fitting with Bayesian inference methods.}
   {We compute SEDs based on the synchrotron self-Compton model for blazar emission using the radiation code \atheva, and use them to train Neural Networks exploring whether these can replace the original computational expensive code.}
   {We find that a Neural Network with Gated Recurrent Unit neurons can effectively replace the \atheva leptonic code for this application, while it can be efficiently coupled with MCMC and nested sampling algorithms for fitting purposes. We demonstrate this through an application to simulated data sets and with an application to observational data. We offer this tool in the community through a public repository.}
   {We present a proof-of-concept application of neural networks to blazar science. This is the first step in a list of future applications involving hadronic processes and even larger parameter spaces.}


   \maketitle
%

\section{Introduction}

Active galactic nuclei (AGN) are known to launch relativistic and collimated plasma outflows, known as jets. While jetted AGN consist a small fraction of the AGN population \citep{2017A&ARv..25....2P}, they have attracted a lot of attention over the past decades because they are the most powerful and persistent sources of non-thermal electromagnetic radiation \citep[for recent reviews, see][]{2019ARA&A..57..467B, 2019NewAR..8701541H}. AGN with jets that are directed toward us are known as blazars. The combination of the small viewing angle with the relativistic motion of the emitting plasma results in strong Doppler beaming, making blazars ideal sources for studying non-thermal radiation processes in jets.

Central to the study of blazars is their spectral energy distribution (SED), which can be thought of as a metric that maps the energy output of the jet across different photon energies, presenting a panoramic view of the radiative processes within the jet. Decoding the SED is pivotal for uncovering the prevailing physical conditions and radiative processes at play in the otherwise out-of-reach regions of extragalactic jets. A significant hurdle in this process, however, is the fitting of the SED itself, which relies on a juxtaposition of the observed energy distributions with model predictions. Radiative models for jet emission are primarily numerical, as they require the solution of a system of stiff coupled partial differential equations (PDEs) describing the evolution with time of the distribution functions of radiating particles and photons~\citep[e.g.][]{DMPR12, AM3, LeHa-Paris, SOPRANO}. Such numerical codes tend to have a high computational complexity, especially in the case of lepto-hadronic models\footnote{Models that account for the radiation processes of relativistic electrons and protons are known as (lepto-)hadronic.}, where each SED computation could last from a few minutes up to an hour (depending on the numerical schemes used by the PDE solver, but also on the approximations made for each operator in the PDE). If such numerical model were to be used in a Markov Chain Monte Carlo (MCMC) algorithm to determine the posterior distributions of the model parameters, the overall execution time could become prohibitively long.  Addressing this computational bottleneck is central to accelerating the pace of discovery in blazar science. 

Although the idea behind a Neural Network (\nn) has been around for more than 50 years \citep{mcculloch1943logical}, it's extended application has been feasible with hardware advantages over the last decade. 
Inspired by recent trends in computational biology \cite{2019NatCo..10.4354W} and similar applications in other disciplines \citep[e.g.][]{2020NanoL..20..329W,2022Symm...14.2482A}, this work investigates the possibility of replacing a numerical model of average computational complexity, i.e. the synchrotron self-Compton\footnote{This is a leptonic model where low-energy synchrotron photons produced by relativistic electrons are up-scattered to high-energies via inverse Compton off the same electron population. } \citep[SSC, e.g.][]{1992ApJ...397L...5M,1996ApJ...461..657B, 1997A&A...320...19M} with a \nn \citep[see][]{2018Heliy...400938A}. 
In pursuit of this objective, the initial phase involves the meticulous assembly of a dataset derived from the numerical model. While the intricate processes of dataset generation are set aside for brevity, it is imperative that this dataset is comprehensive and uniformly spans the parameter space. Furthermore, this compilation should be both ample in size to enable rigorous training and analysis, and efficiently constructed to conserve time and resources. With a robust dataset in hand, the exploration of various \nn  architectures takes center stage. 

The realm of machine learning (\ml) is characterized by its blend of scientific rigor and artistic intuition; there exists no monolithic model that stands as the unequivocal solution. Each model is a unique amalgamation of structural nuances and algorithmic intricacies, tailored to address specific aspects of complex problems. Consequently, we will investigate a suite of \nn configurations to identify those offering optimal performance and accuracy in the context of the defined problem space.
 
This paper is structured as follows. In Sec.~\ref{sec:methods} we outline the methodology used to construct a model based on a \nn, which includes the use of a numerical code to construct a dataset and the various \nn tuning and evaluation tests. In Sec.~\ref{sec:results} we present a few scenarios where the \nn trained model is coupled with a Bayesian interface to fit simulated data and actual blazar observations. We also provide notes on the efficiency of our approach compared to numerical codes. Finally, we present the conclusions of this work and some future aspects in Sec.~\ref{sec:conclusions}.

\begin{figure}
\centering
\includegraphics[width=0.49\textwidth]{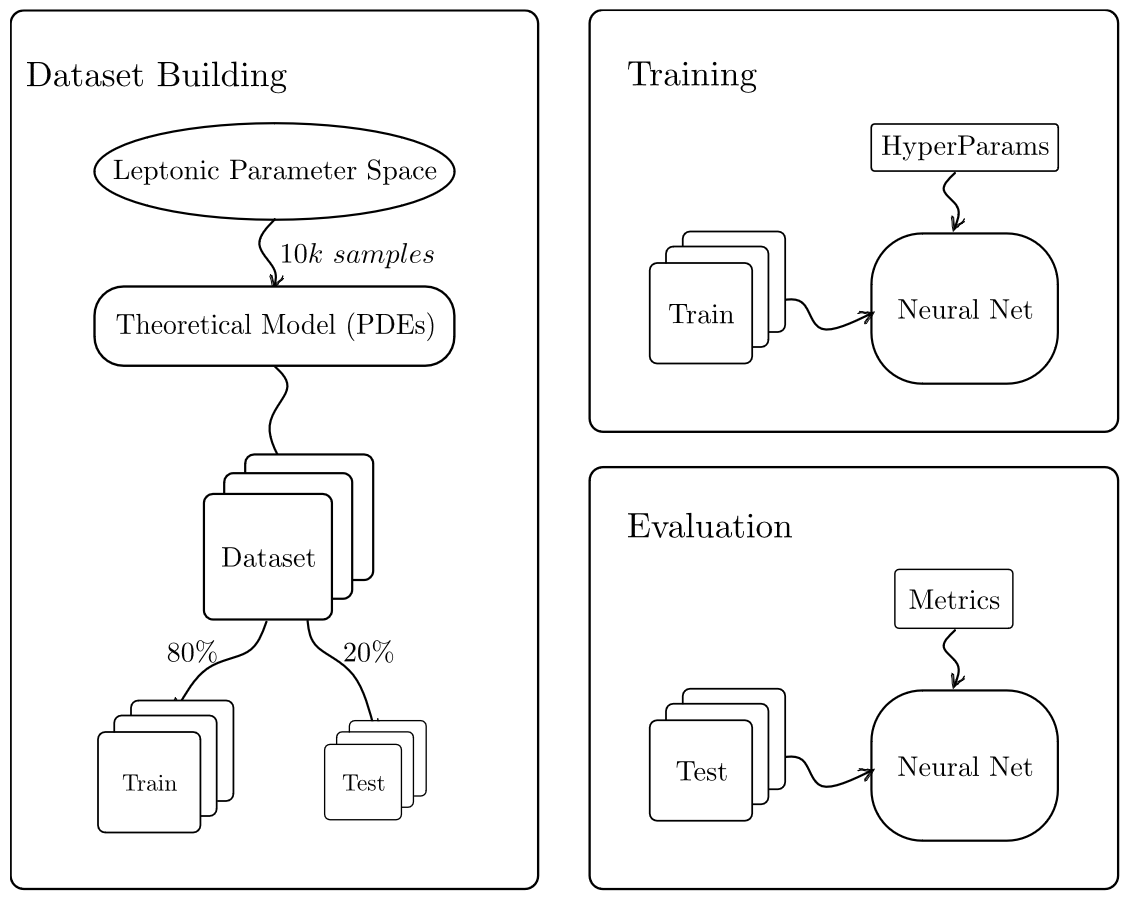}
\caption{Implementation flow chart describing our methodology.}
\label{fig1}       
\end{figure} 

\section{Methods}\label{sec:methods}
Our approach, which is schematically shown in Fig.~\ref{fig1}, is comprised of the following steps: (i) we built a  database of blazar SED models, (ii) we train the \nn using available datasets, and (iii) we evaluate our results based on datasets not used in the training. In the following paragraphs we elaborate on the above steps.

\subsection{Generation of dataset}
To generate blazar SEDs we use the radiative transfer code \atheva \citep{1995A&A...295..613M,2012A&A...546A.120D} that solves a system of coupled PDEs describing the evolution of relativistic particle distributions and photons contained in a non-thermal emitting source. 
For this project we use the leptonic module of the code that describes a magnetized source containing relativistic electron-positron pairs and photons. Electrons are being injected into the source at a fixed rate (or injection luminosity), and escape on an energy-independent timescale that is equal to the light-crossing time of the source. Photons are being produced via electron synchrotron radiation and scattered to higher energies through inverse Compton scattering, while they escape on the light-crossing time from the source\footnote{Additional processes, like photon-photon pair production and synchrotron self-absorption, that may affect the lowest and highest energy parts of the photon spectrum are not included in this proof-of-concept work. Trained \nns with these procedures will be gradually added to the GitHub repository of the project.}. For a detailed description of the modeling of each physical process, we refer the reader to \citet{1995A&A...295..613M, 2012A&A...546A.120D}). 

The SSC model parameters, which are used as an input to \atheva, are:
\begin{itemize}
    \item the characteristic size of the emitting region $R$ (i.e., its radius assuming a spherical source in the jet rest frame),
    \item the magnetic field strength $B$ (measured in the jet rest frame),
    \item the electron injection compactness $\ell_e$, which is a dimensionless measure of the power injected into relativistic electrons (defined as $\ell_e = \sigma_T L_{e}/(4 \pi R m_e c^3)$),
    \item the power-law slope $p$ of the electron distribution at injection (i.e., $dN_e / dt d\gamma_e \propto \gamma_e^{-p}$), and
    \item the Lorentz factor range of radiating electrons, $[\gamma_{min}, \gamma_{max}]$, also at injection.
\end{itemize}

For the generation of the dataset we use parameter values that are motivated by  SSC models of blazar emission~\citep[e.g.][]{2008MNRAS.385..283C}. For instance, we take the lower bound of the electron distribution to range between $10^{0.1}$ and $10^4$. To avoid very narrow electron distributions, or cases where $\gamma_{max}<\gamma_{min}$ that will lead to code crashing, we set the upper cutoff of the distribution to lie between $10^2 \gamma_{min}$ and $\min(10^8, \gamma_{H})$; here $\gamma_H = eBR/(m_e c^2)$ is an upper limit imposed by the Hillas condition \citep{1984ARA&A..22..425H}, which ensures magnetic confinement of the most energetic electrons in the source. Each variable is randomly drawn from a uniform  distribution in logarithmic space (expect for the slope $p$) --  see Table \ref{tab:paramrange}. Given the theoretically known dependence of the synchrotron/SSC fluxes and the characteristic photon energies on model parameters, like $R$ and $B$ \citep[see e.g.][]{1997A&A...320...19M, 2008ApJ...686..181F}, the selected ranges are wide enough to produce a diverse dataset, which is crucial for the \nn training (see next subsections). 

Another parameter that should be taken into account for the SED fitting is the Doppler factor $\delta$ of the emitting region, which is used to transform the photon spectra from the jet comoving frame to the observer's frame. However, this is a nuisance parameter for the dataset creation, since the Doppler boosting can be applied at a later stage to transform the comoving photon spectra (unprimed quantities) to the observer's frame, i.e. $\nu_{\rm obs} = \delta \nu$ and $L_{\nu, \rm obs} = \delta^3 L_{\nu}$ (here, $\nu$ is the photon frequency and $L_\nu$ is the specific luminosity).

To generate the dataset we sample the space of leptonic parameters and feed each set to \atheva. The system evolves for $5~R/c$, which is sufficiently long for the establishment of a steady state (equilibrium solution of the PDEs). The parameters along with the photon PDE solution are concatenated to form the entry of each dataset. All SEDs were computed at the same 500 grid points of photon energies, logarithmically spaced between $10^{-15}$  and $10^{10}$ (in units of $m_{\rm e}c^2$). The SED fluxes returned by \atheva at each grid point ($\nu F_{\nu}$ in code units)  were re-normalized using Min-Max scaling between the values of 0 and 1, before being used for the training. Normalization of inputs is standard practice \citep{chris1995neural}. In total, we created a dataset of 10,000 samples, for the range of parameters listed in Table \ref{tab:paramrange}, within $\sim$500 CPU hours (i.e. about 3~min per model) on a desktop computer (AMD Ryzen 5950X). The generated dataset is split into 80\% for training, 10\% for validation and 10\% for testing.

\begin{table}
\centering
\caption{Input parameter ranges used for the generation of training and validation datasets with the \atheva code. $U()$ stands for a uniform distribution.} 
\begin{tabular}{lc}
\hline
\textbf{Parameter [unit]} & \textbf{Range}  \\
        \hline \hline
        $\log R$  [cm]   & $U(14, 17)$ \\
        $\log B$  [G]   & $U(-2, 2)$ \\
        $\log \gamma_{min}$ & $U(0.1, 4)$   \\
        $\log \gamma_{max}$ & $U(\log 10^2\gamma_{min}, \min\left(8, \log \gamma_{H}\right))$  \\
        $\log \ell_e$ & $U(-5, -1)$  \\
        $p$     & $U(1.5, 3)$  \\
        \hline
\end{tabular}
\label{tab:paramrange}  
\end{table}

\subsection{Training of the \nns}
Once a dataset has been obtained, it is essential to select the appropriate type of \nn neuron and its corresponding structure. The prediction of the SED can be perceived as a regression problem, making the simple artificial neuron an apparent choice. However, it is also possible to approach the bins of the SED as sequential data.
In this case, all three types of recurrent neurons become potential options for constructing the \nn.
We employed several \nn topologies that varied with respect to the following parameters: number of hidden layers,  number of neurons per hidden layer, and three types of neurons: an artificial \nn with a deep stack of hidden layers, i.e. Artificial Neural Network (ANN), a Gated Recurrent Unit \citep[GRU,][]{cho2014properties} and a Long Short-Term Memory \citep[LSTM,][]{hochreiter1997long}.

Currently, there are a few \ml frameworks available for several programming languages that can be used for practical applications. Python programming language is a very popular choice in the astrophysics community, and as a result there are a lot of scientific software packages available for use. The major \ml packages available in Python are: {\tt Tensorflow} created by Google Brain \citep{tensorflow2015-whitepaper}, {\tt PyTorch} developed by Facebook's AI Research lab \citep{NEURIPS2019_9015}, {\tt scikit-learn} built by David Cournapeau \citep{scikit-learn}, {\tt Theano} developed by the Montreal Institute for Learning Algorithms \citep{2016arXiv160502688T}.
{\tt Tensorflow} and {\tt Theano} use {\tt Keras} \citep{chollet2015keras}, an open-source software library that provides a Python interface for \nns. It was developed with a focus on enabling fast experimentation and prototyping, being both user-friendly and modular. In this work we opted for {\tt Tensorflow}. 

We trained the 3 \nns using the 10k dataset; the hyper-parameters and complexity of each \nn are presented in Table \ref{tab:tuner_params}.
An additional necessary parameter is the number of epochs, which refers to the number of training cycles. The appropriate number of epochs must be carefully selected for each network to address the issue of over-fitting.
To determine the optimal number of epochs for each network, an initial training phase is conducted with a substantially large number of epochs (set at 3,000), since experience demonstrated that over-fitting, if it occurs, typically starts after 200 epochs. Finally, one can choose a loss function other than the typical mean-squared error (MSE). Since our networks handle normalized data it was decided to use instead the mean squared logarithmic error (MSLE) as a loss function.
The training process of these networks is observed in real-time using Tensorboard. This enables the monitoring of whether a network is over-fitting or if it has achieved a satisfactory loss error, which is $\mathcal{O}(10^{-4})$. Whichever of these conditions is met first dictates the stopping point. The epoch at which training is stopped is noted, and then the network undergoes retraining for this specific number of epochs.

\begin{table} 
    \centering
    \caption{Hyper-parameters for the \nns.} 
    \label{tab:tuner_params}
    \begin{threeparttable}
    \begin{tabular}{l | ccc}
        \hline
        \diagbox{Parameter}{Cell Type} & ANN & GRU & LSTM\\
        \hline \hline
        Hidden Layers & 6 & 6 & 6 \\
        Neurons per Layer & 184 & 184 & 184\\
        Learning Rate ($\times10^{-4}$)& $1.7$ & $1.7$ & $1.7$\\
        Batch Size & 32 & 32 & 32 \\
        \nn Complexity\tnote{*} & 300k& 1.4M& 1.9M\\
        \hline
    \end{tabular}
    \begin{tablenotes}
    \item[*]Total number of model parameters in thousands (k) or millions (M).
    \end{tablenotes}
    \end{threeparttable}
\end{table}

\subsection{Evaluation of results}

\begin{table} 
    \centering
    \caption{Evaluation of results.} 
    \label{tab:valid}
    \begin{threeparttable}
    \begin{tabular}{p{0.12\columnwidth}p{0.24\columnwidth}p{0.24\columnwidth}p{0.2\columnwidth}}
        \hline
        & \multicolumn{3}{c}{Cell Type}\\
        \hline
         & ANN & GRU & LSTM\\
        \hline \hline 
        Metric & \multicolumn{3}{l}{mean score\tnote{*}} \\
        \hline
        R$^2$  & 0.902 & 0.985 & 0.989 \\
        MSE  & 1.84 & 0.23 & 0.20 \\
        MAE  & 0.89 & 0.13 & 0.13 \\
        KS   & 0.256 & 0.070 & 0.068\\
        DTW  & 8.86 & 3.12 & 4.37 \\
        \hline\hline
         & \multicolumn{3}{l}{Ranking percentages (\%)\tnote{**}} \\
         \hline
        MSE  & 0/1.3/98  &  42/56/1 &  57/42/0 \\
        MAE  & 0/0 /100  &  79/ 21/0 &  21/79/0 \\
        KS   & 0.5/2.9/97  & 54/44 /1.8  &  46/53/1.6 \\
        DTW  & 0.00/1.2/99 & 87/12/0.7 &  13/86/0.4  \\
        \hline
    \end{tabular}
    \begin{tablenotes}
    \item[*] Median metric value for each \nn based on the test sample. 
    \item[**] Percentage of times where each \nn ranked 1$^{\rm st}$/2$^{\rm nd}$/3$^{\rm rd}$ in their between comparison for the test sample. 
    \end{tablenotes}
    \end{threeparttable}
\end{table}

\begin{figure*}
\centering
\xincludegraphics[scale=0.31,label=(a),pos=n]{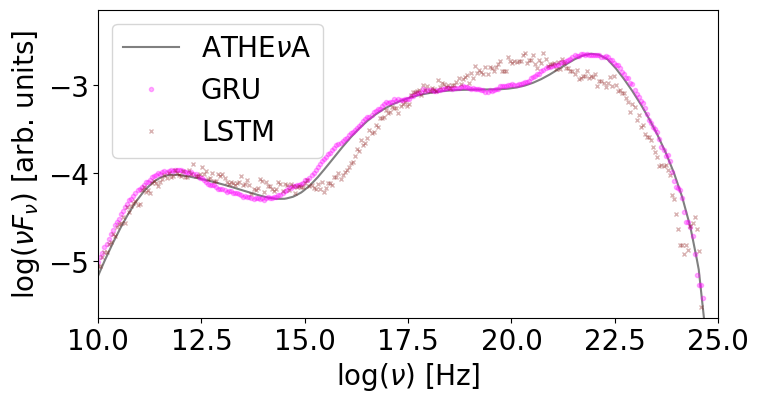}
\xincludegraphics[scale=0.31,label=(b)]{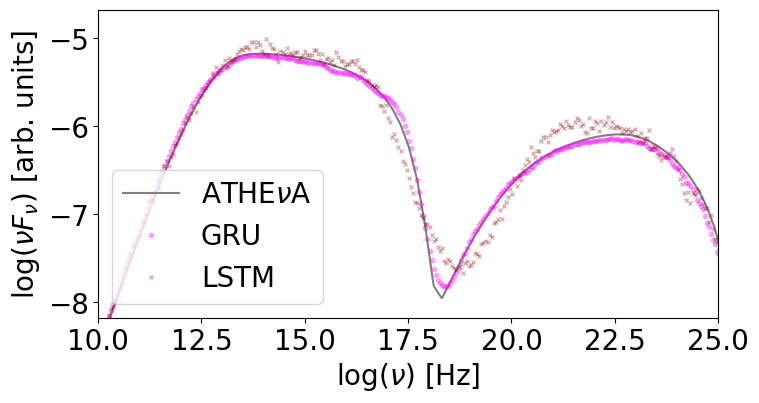}
\xincludegraphics[scale=0.31,label=(c)]{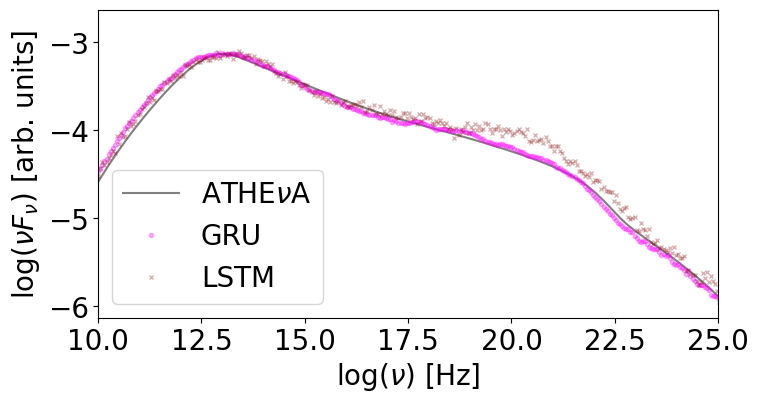}
\caption{Visual comparison of GRU (magenta points) and LSTM (maroon points) \nns on top of the \atheva SED. Both \nn generated models capture the overall trends, however GRU offers a better description of the \atheva SED with less scatter.} 
\label{fig:SED_GRU_LSTM}       
\end{figure*}

\begin{figure*}
\centering
\xincludegraphics[scale=0.33,label=(a)]{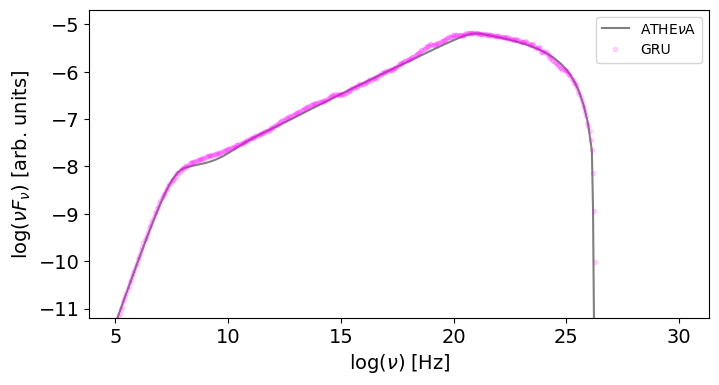}
\xincludegraphics[scale=0.33,label=(b)]{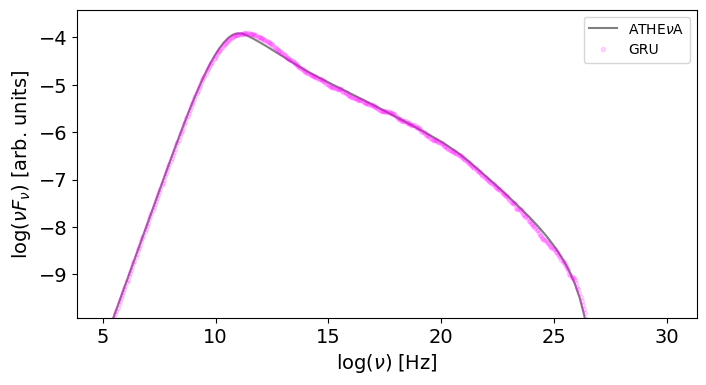}
\xincludegraphics[scale=0.33,label=(c)]{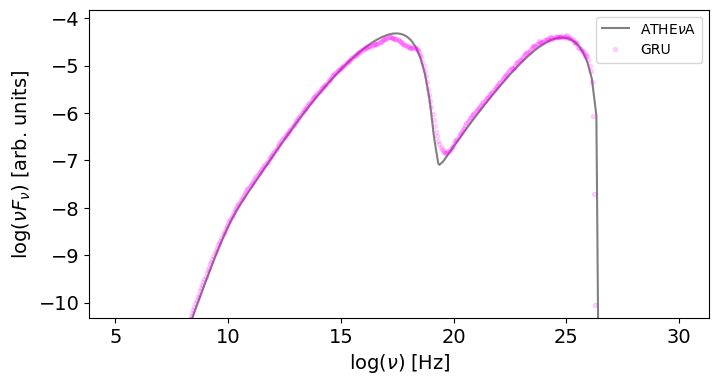}
\\
\xincludegraphics[scale=0.33,label=(d)]{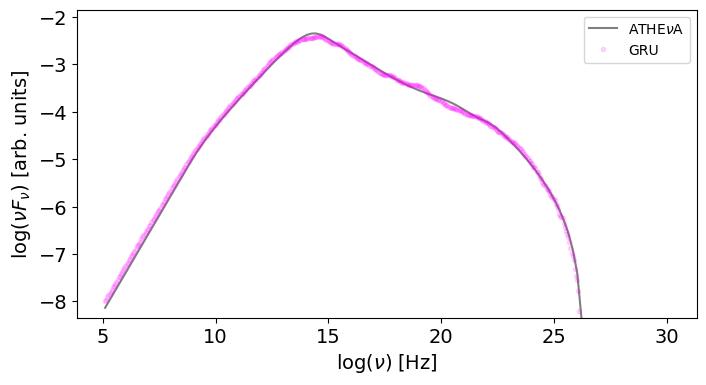}
\xincludegraphics[scale=0.33,label=(e)]{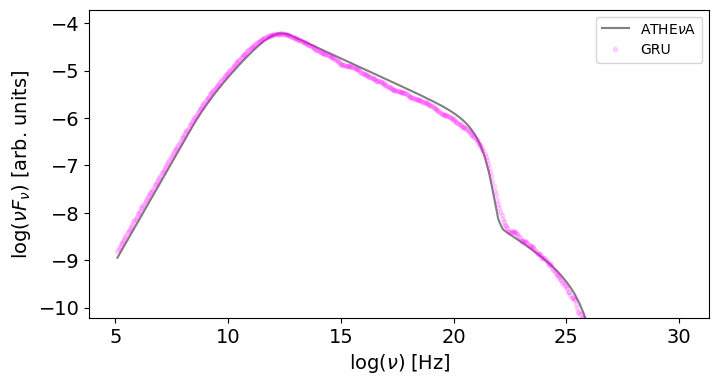}
\xincludegraphics[scale=0.33,label=(f)]{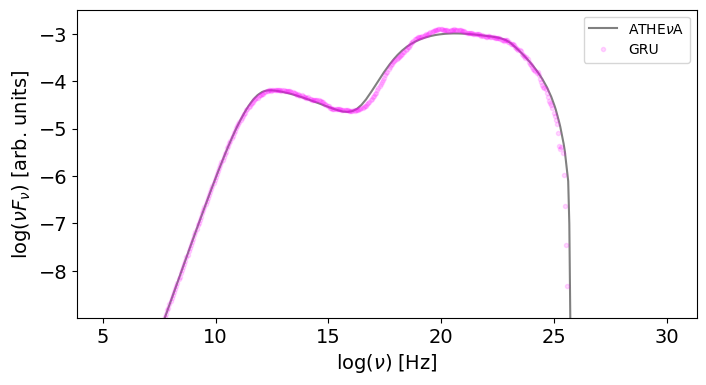}
\\
\xincludegraphics[scale=0.33,label=(g)]{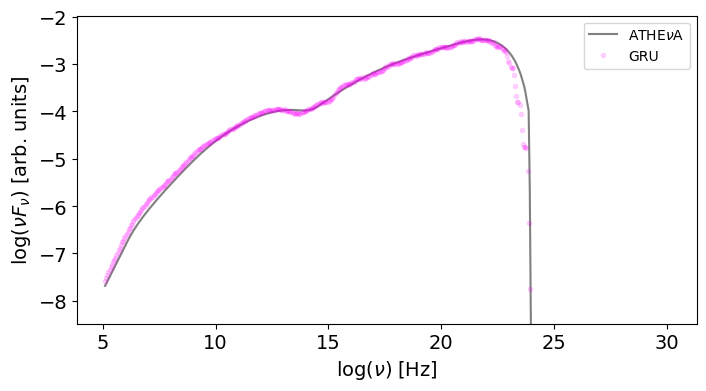}
\xincludegraphics[scale=0.33,label=(h)]{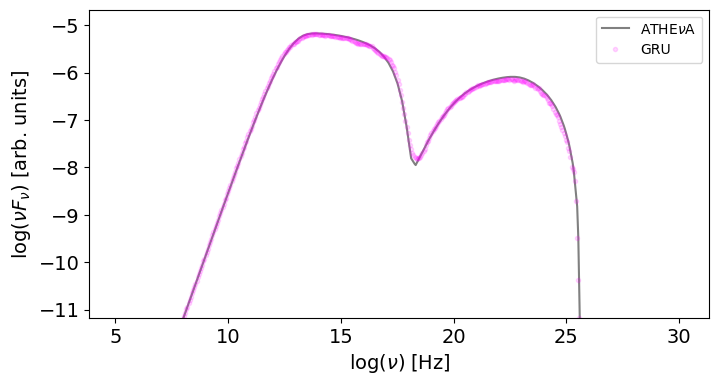}
\xincludegraphics[scale=0.33,label=(i)]{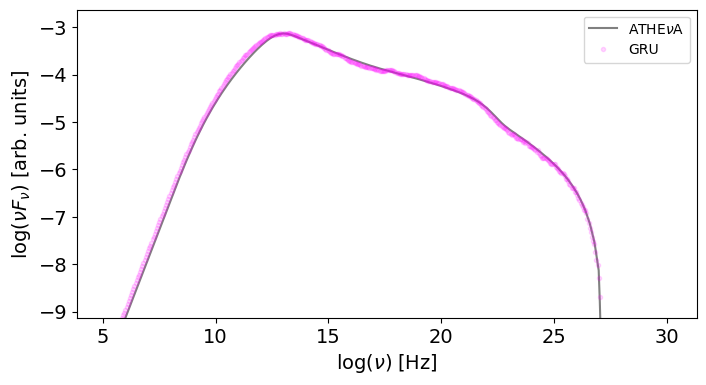}
\\
\xincludegraphics[scale=0.33,label=\ \ \ \ \ \ \ \ \ \ (j)]{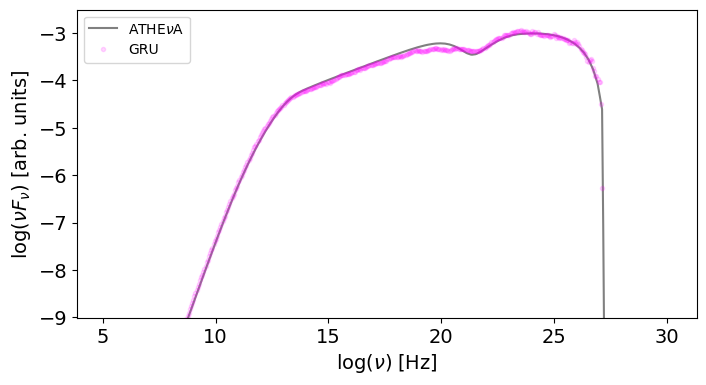}
\xincludegraphics[scale=0.33,label=(h)]{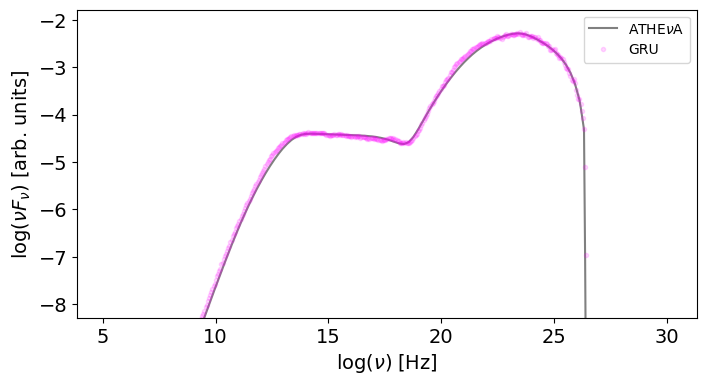}
\xincludegraphics[scale=0.33,label=(k)]{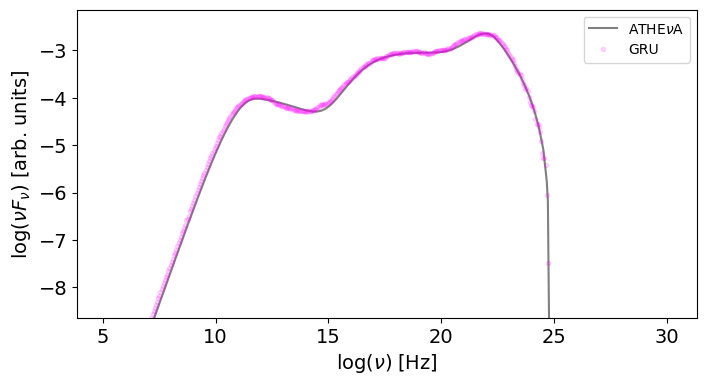}
\\
\xincludegraphics[scale=0.33,label=(l)]{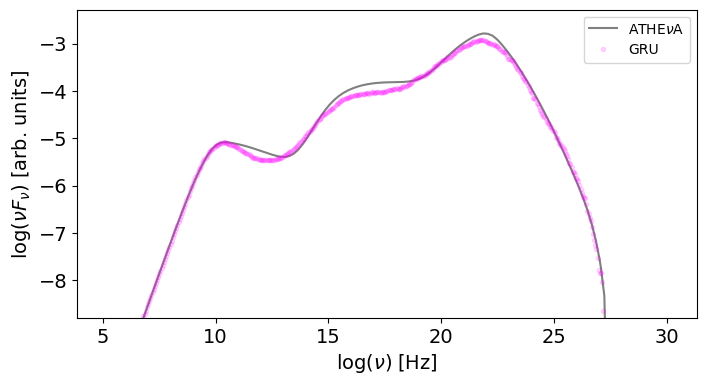}
\xincludegraphics[scale=0.33,label=(m)]{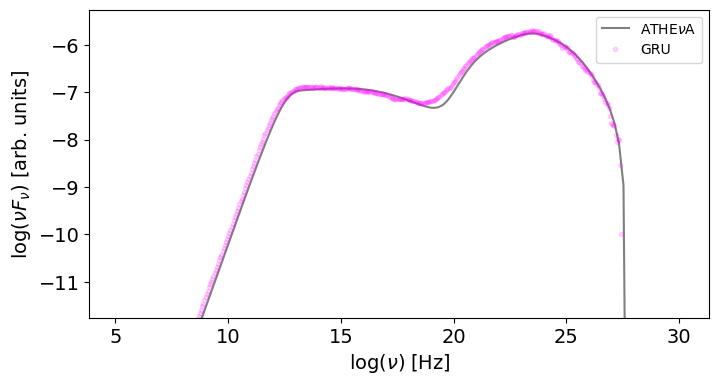}
\xincludegraphics[scale=0.33,label=(n)]{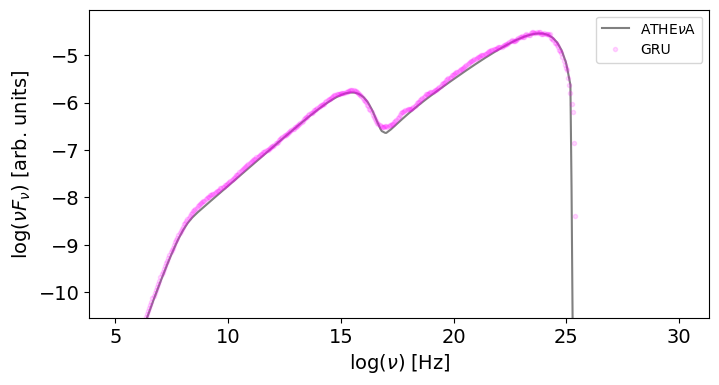}
\\
\xincludegraphics[scale=0.33,label=(o)]{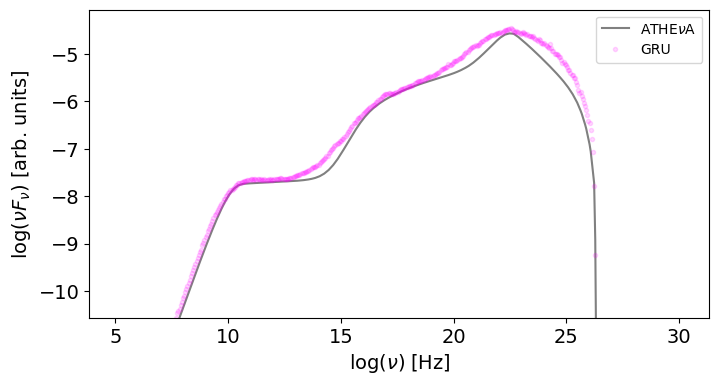}
\xincludegraphics[scale=0.33,label=(p)]{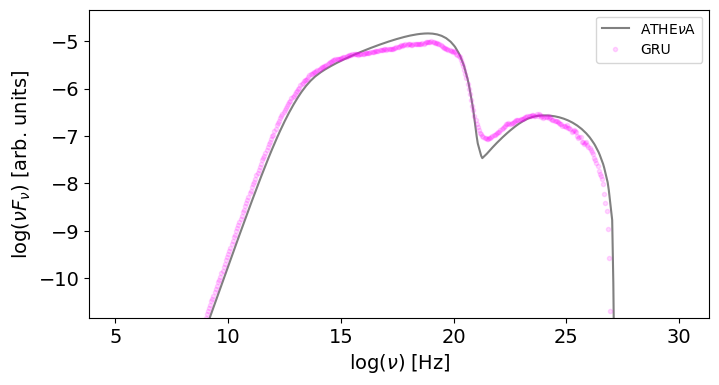}
\xincludegraphics[scale=0.33,label=(q)]{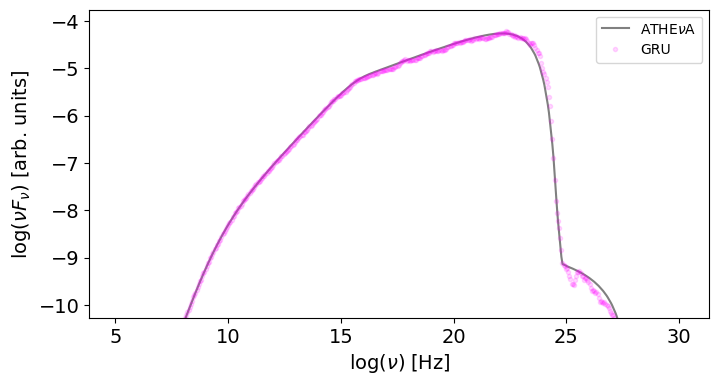}
\\
\xincludegraphics[scale=0.33,label=(r)]{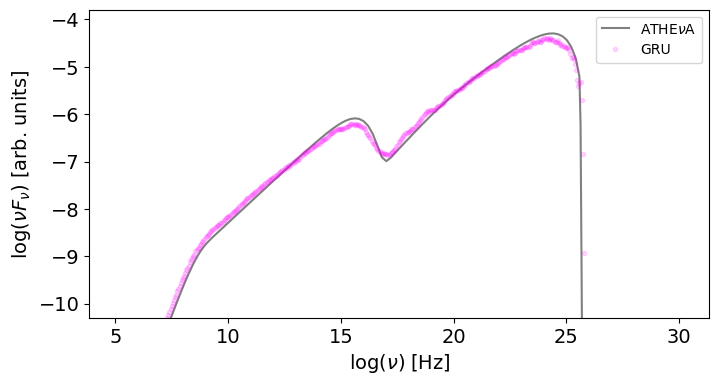}
\xincludegraphics[scale=0.33,label=(s)]{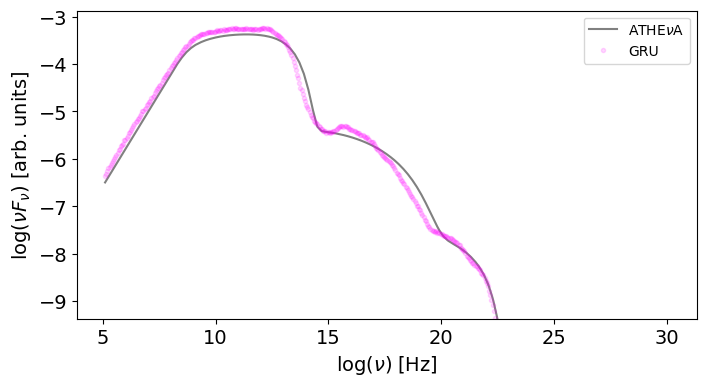}
\xincludegraphics[scale=0.33,label=(t)]{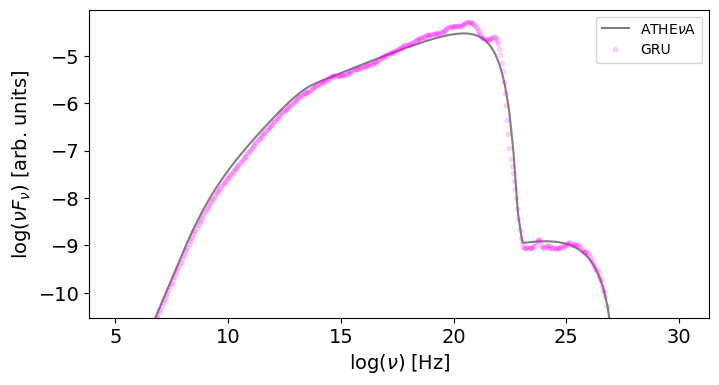}
\caption{Examples of SEDs from the test dataset (i.e., not used for the training). Predictions of the trained GRU \nn are overplotted (colored markers) to the numerical models created with \atheva (solid lines).}
\label{fig:Test_spec}       
\end{figure*}

 After training our 3 \nns we proceed to evaluation of the models by comparing the \nn predictions to the SEDs from the test sample.
We first use standard metrics, namely the R$^2$ score, the MSE, the mean-squared absolute error (MAE), the Kolmogorov-Smirnov (KS) criterion, and  Dynamic Time Warping (DTW), to compare the performance of the trained \nns. 
In terms of R$^2$ the GRU and LSTM \nns have significantly better values than the ANN network. The ranking results, which are listed in Table \ref{tab:valid}, show that the GRU is ranked first among the three by three metrics, while the ANN is ranked third by all criteria. We also visually inspect the test dataset, where we compare the prediction of the \nns against the result of \atheva -- see Fig.~\ref{fig:Test_spec}. From this qualitative test, it became clear that ANN failed to describe the SED, suggesting that more complexity has to be added to the network. Meanwhile, both GRU and LSTM delivered acceptable solutions (based on the eyeball test). The main difference was that GRU offered more smooth solutions for the SED, while the LSTM predictions had noticeably small amplitude scatter across neighboring values. In combination with the evaluation results listed in Table \ref{tab:valid}, we select the GRU \nn for the applications that follow.

In Fig.~\ref{fig:Test_spec} we present random examples drawn from the test dataset that highlight the wide variety of SEDs (in terms of flux and shape) reproduced by the GRU \nn, including spectra that are not typical of blazars. For instance, there are cases without a clear two-hump morphology due to the smooth superposition of the synchrotron and SSC components (see e.g. panels a and b), or due to a very extended synchrotron spectrum (panel q), which reminds more of spectra of pulsar wind nebulae \citep[see e.g.,][for Crab nebula]{universe7110448}. Moreover, the sample contains spectra produced in the so-called inverse Compton catastrophe limit \citep{1994ApJ...426...51R, 2015MNRAS.452.3226P}; these SEDs consist of three distinct components attributed to synchrotron, first-order and second-order SSC emission, with the latter carrying most of the bolometric power of the source (see panels l and o).
In most cases, the trained \nn describes accurately the numerical model (e.g. panels a, h and i). There are also cases where the \nn prediction misses the details of the model, but still captures the general trend (e.g. panels t, o, and p). We will show in the following section, that such differences are not crucial for the interpretation of observed spectra in the context of fitting with Bayesian methods.

\begin{figure*}
\centering
\xincludegraphics[scale=0.41,label=Case 1]
{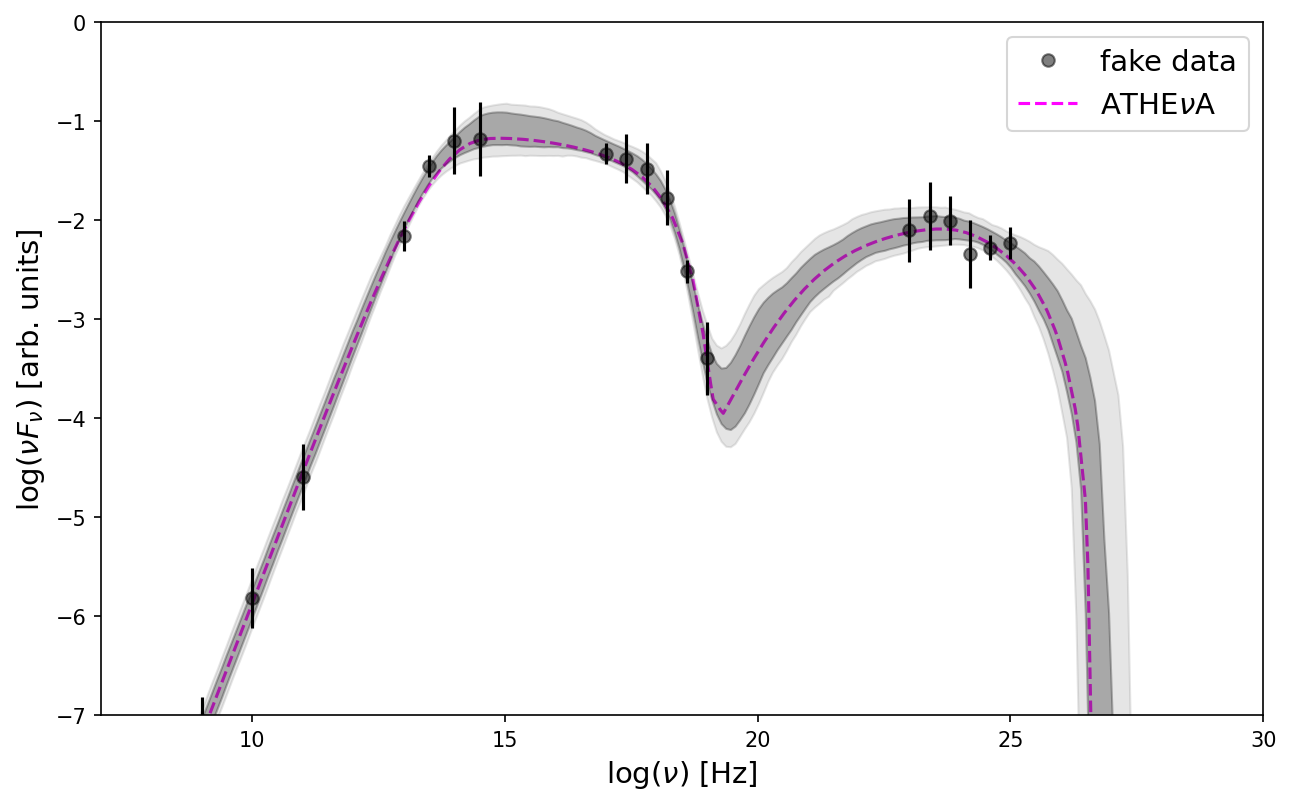}
\xincludegraphics[scale=0.41,label=Case 2]
{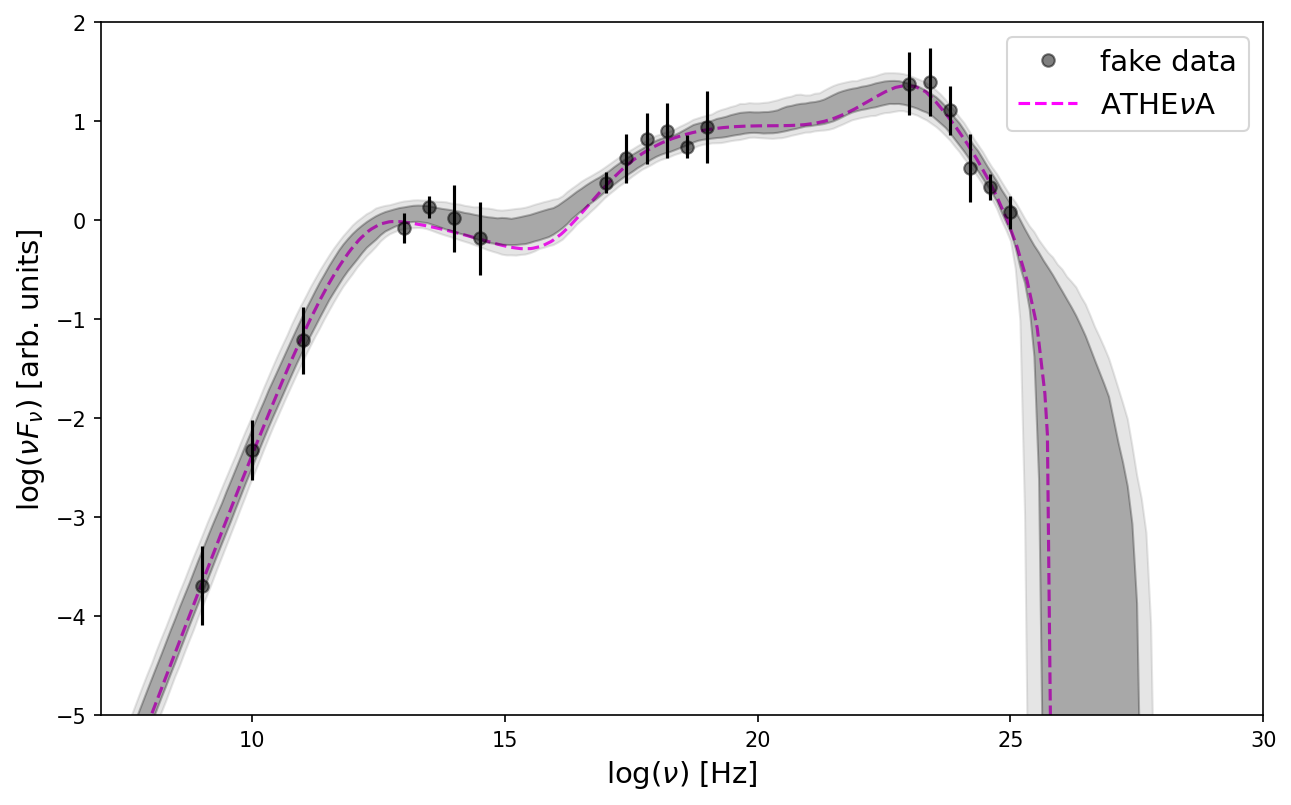}
\caption{Simulated blazar SEDs fitted with the GRU \nn generated model. Simulated data are shown with markers, and with dashed magenta line is plotted the SED from the \atheva code. Dark- and light-shaded areas denote the range of GRU \nn models with parameters from the 50\% and 90\% posterior distributions, respectively.} 
\label{fig:fit_SED}       
\end{figure*}

\begin{table*}
\caption{Values from GRU \nn fitting to simulated SEDs with {\tt emcee}.} 
\centering
\begin{tabular}{lcccc}
\hline
& \multicolumn{2}{c}{\textbf{Case 1}} & \multicolumn{2}{c}{\textbf{Case 2}} \\
        \hline \hline
\textbf{Parameter [unit]} & True & Fit& True & Fit\\   
\hline
        $\log R$  [cm]  & 16.60& $16.04^{+0.61}_{-0.97}$ & 15.42& $14.78^{+0.66}_{-0.53}$\\
        $\log B$  [G]   & -0.34& $-0.49^{+0.36}_{-0.37}$& -0.12& $0.24^{+0.38}_{-0.38}$\\
        $\log \gamma_{min}$ & 3.37& $3.13^{+0.35}_{-0.24}$& 2.31&   $2.20^{+0.31}_{-0.26}$\\
        $\log \gamma_{max}$ & 5.55& $5.51^{+0.25}_{-0.19}$& 4.76&  $5.57^{+1.07}_{-0.98}$\\
        $\log \ell_e$ &-4.57 & $-5.08^{+0.40}_{-0.49}$& -1.82& $-1.83^{+0.46}_{-0.29}$\\
        $p$     & 2.20& $2.76^{+0.17}_{-0.34}$& 2.62&  $2.64^{+0.20}_{-0.22}$\\
        $\delta$ & 1& $1.3^{+0.2}_{-0.2}$& 1& $1.00^{+0.16}_{-0.14}$\\
        \hline
\end{tabular}
\label{tab:fake_fit}  
\end{table*}

\section{Results}\label{sec:results}

\subsection{Simulated SEDs}

A fundamental question is whether a trained \nn can be used to recover the physical parameters corresponding to an observed SED, provided one knows the ``correct'' answer. The only way to test this is to use simulated SEDs based on an SSC model with known parameters,  which are then fitted by the \nn; the fitted values can be then compared to the true values of the SSC model to address the question of whether the \nn can constrain the parameters with sufficient accuracy. 

To create a simulated SED we select 25 points of an SSC model computed with \atheva from a wide frequency range and add Gaussian noise with 0.1 standard deviation to the logarithmic flux values. We then attribute an error to each flux point between 0.1 and 0.4 in logarithm to mimic statistical uncertainties in the measurement. As blazars are observed in specific energy bands, we select points that broadly correspond to radio, UV, X-ray and $\gamma$-ray bands. By creating an incomplete SED with gaps in regions where the solution may have large derivatives (e.g., spectral cutoffs), we increase the difficulty for the \nn fitting. We also note that the simulated flux points may overshoot the baseline model of \atheva in certain frequency bands due to the randomness of the simulation and the small number of selected points. This mimics observational effects related to non-simultaneous observations and intrinsic source variability.

We used the trained GRU \nn to fit the data in Python with {\tt emcee}\footnote{A Python implementation of the Affine invariant Markov chain Monte Carlo (MCMC) ensemble sampler.} \citep{2013PASP..125..306F}. For constructing the log-likelihood function and fitting the data we followed an often-adopted methodology in similar problems \citep[e.g.][]{2023MNRAS.520..281K,2023arXiv230806174S}. 
We also added a term $\ln{f}$ as a free parameter to account for the systematic scatter and noise of the simulated data and the \nn model. By including $\ln{f}$ we get an excess variance compared to statistical uncertainties, i.e.
$\sigma_{\rm tot, \rm i}^2 = \sigma_{\rm i}^2 + e^{2\ln{f}}$, where $\sigma_{\rm i}$ is the error of individual flux points.

Here we present results for two characteristic cases. The simulated SEDs (markers) together with the corresponding SED from the \atheva code (dashed line) and the fitted models (shaded regions) are shown in Fig. \ref{fig:fit_SED}. The corner plots of the posterior distributions are presented in Figs. \ref{fig:corner-fake12} and  \ref{fig:corner-fake27}.
The \nn model is able to successfully fit the simulated data and capture the overall shape of the true model. In terms of parameters estimations all parameters were consistent within 2$\sigma$ with the original values used for simulating the data sets (see Table \ref{tab:fake_fit}).

\subsection{Observed SEDs}\label{sec:applications}

\begin{figure*}
\centering
\includegraphics[width=0.49\textwidth]{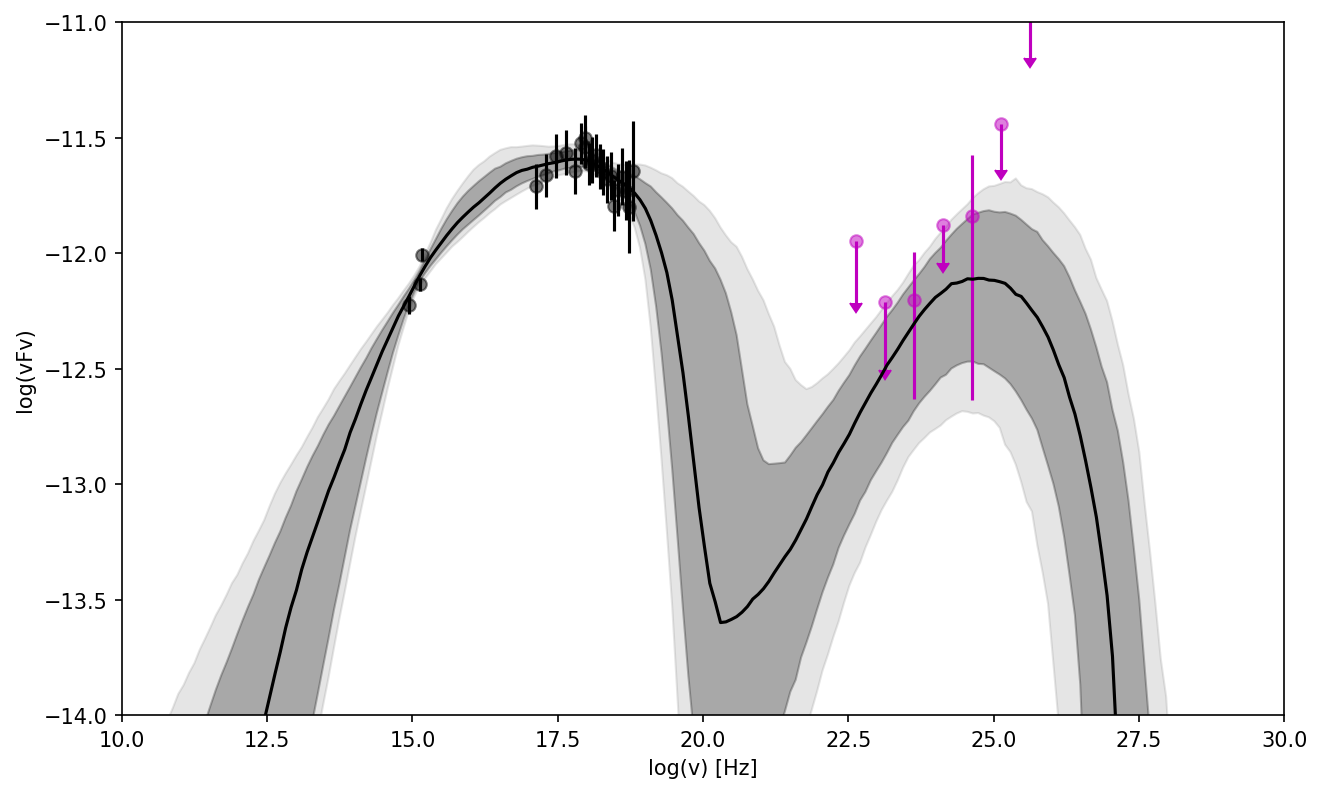}
\includegraphics[width=0.49\textwidth]{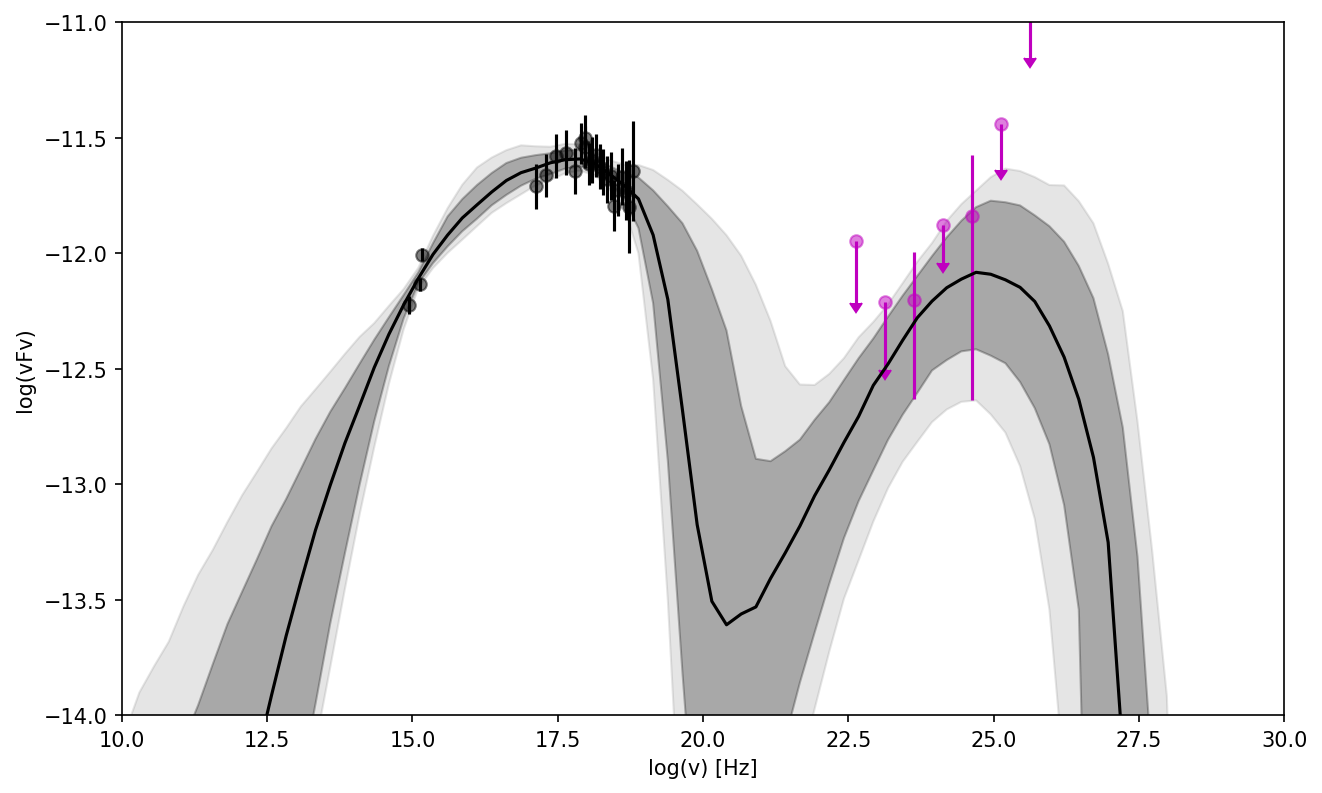}
\caption{Observed SED of \hsp \,  (colored markers) fitted with the GRU \nn generated model, with {\tt emcee} (left) and {\tt UltraNest} (right). Dark- and light-shaded areas denote respectively the 50\% and 90\% confidence regions of the fitted models.} 
\label{fig:fit_SED_EXT}       
\end{figure*} 

We demonstrate the capabilities of the trained \nn model to fit a real blazar SED  with {\tt emcee} and {\tt UltraNest} \citep{2021JOSS....6.3001B}, which employs nested sampling to scan multi-modal posterior distributions. The latter allows us to better explore possible degeneracies in multi-parameter problems.

For this application, we choose \hsp \,  \citep{Giommi2020, 2020ApJ...902...29P}, which is a BL Lac object at redshift $z=0.557$ \citep{10.1093/mnrasl/slaa056, 2020ApJ...902...29P} that belongs to the rare class of extreme blazars \citep{Biteau2020}. This blazar has been detected in GeV $\gamma$-rays by the \fermi \, Large Area Telescope (LAT) and is part of the  4FGL catalog \citep{2020ApJS..247...33A}. \hsp \, is an ideal test-bed because it combines measurements with small errors at lower energies with several upper limits in the $\gamma$-rays. 
Another reason for our source selection was that this target was also recently modeled with an {\tt emcee} adaptation of the leptonic module of \lehamoc\footnote{\url{https://github.com/mariapetro/LeHaMoC}}, a Python-based lepto-hadronic code that offers a factor of $\sim 30-100$ speed improvement compared to \atheva, while producing consistent results \citep{2023arXiv230806174S}. Therefore, we can compare the efficiency and accuracy of SED fitting performed with a fast leptonic code and a trained \nn.

We used the same dataset as \citet{2023arXiv230806174S}. The quasi-simultaneous observations in the UV, soft X-rays, and hard X-rays provide a detailed picture of the low-energy part of the spectrum. On the contrary, the high-energy part of the spectrum is less constrained observationally.
For fitting the data and constructing the log-likelihood function we followed \citet{2023arXiv230806174S}, taking into account asymmetric errors and upper limits in the \textit{Fermi} data. For our application we defined uniform priors to cover all the parameter space used to build the datasets, and a Doppler boost factor (in logarithm) in the range [0,3]. We run {\tt emcee} with 48 walkers and 10,000 steps each. We started with a guess value that resembles the general spectral shape and discard the first 1,000 steps of each chain as burn-in. We also ran {\tt UltraNest} with 400 live points, a maximum number improvement loops of 3 and a target evidence uncertainty of 0.5.

The results of the fits are presented in Fig.~\ref{fig:fit_SED_EXT}. We find no noticeable differences between the {\tt emcee} and {\tt UltraNest} results, demonstrating that the SSC model has no multimodal posterior distributions (these are presented in Figs.~\ref{fig:corner-ssc} and \ref{fig:corner-ssc-ultra}). When compared to the fitting results obtained with a leptonic code, the posterior distributions of parameters obtained by the GRU \nn and \lehamoc  \citep{2023arXiv230806174S} are identical in terms of median values, standard deviation and overall distribution shape as shown in the corner plot of Fig.~\ref{fig:corner-ssc}. Moreover, the physical processes that were neglected in the \nn training (i.e. synchrotron self-absorption and $\gamma \gamma$ pair production) turn out to be negligible for this application, since both processes were included in the fitting with \lehamoc.

\subsection{Efficiency of \nns}
We briefly comment on the efficiency of the \nn generated model in comparison to the other numerical codes used in our tests. 

The \atheva code, which was first presented in the mid-nineties \citep{1995A&A...295..613M}, did not target fitting large samples of data. As a result, the numerical scheme adopted in \atheva is not optimized for speed but for accuracy \citep[see also][for comparison of \atheva against other proprietary codes]{2022icrc.confE.979C}. The typical execution time of \atheva for an SSC model is about 1-4~min. Moreover, the code may fail for certain input parameters (when these lead to very large derivatives in the PDEs), which cannot be determined a priori. For instance, about 4\% of the cases failed when creating the dataset for the \nn training. \lehamoc, on the other hand, is more flexible for the generation of big datasets and its use by fitting algorithms \citep{2023arXiv230806174S}: it delivers a factor of $\sim 30-100$ improvement in speed compared to \atheva, and does not crash regardless of the input parameters. When combined with the parallelization option of {\tt emcee} \lehamoc can fit the SED of \hsp \, within 1 day (using 16/32 CPU cores/threads). On the contrary, fitting of \hsp \, using the GRU \nn model was completed in approximately 20~min on a single core, which is about 1,000 faster compared to \lehamoc. 

Therefore, implementing a \nn like GRU offers a huge improvement in computational time as it makes MCMC fitting with {\tt emcee} in a matter of minutes. When combined with nested sampling algorithms like {\tt UltraNest}, the efficiency of the sampling can drop significantly, as it is common for high-dimension problems.
For our application to \hsp \, about 20 million GRU \nn models were evaluated to search all the parameter space (see Table \ref{tab:paramrange}) within 21~hr. In Table \ref{tab:efficieny} we list execution times for the various cases discussed in this work. All computational tasks were performed with the same hardware (AMD Ryzen 5950X). Quoted times were scaled to a single CPU core when parallel computing was used. Times may vary depending on the hardware used, and therefore the listed values in Table \ref{tab:efficieny} are of comparative value.

\begin{table} 
    \centering
    \caption{Typical CPU times for SSC computations.} 
    \label{tab:efficieny}
    \begin{threeparttable}
    \begin{tabular}{lccc}
        \hline
         & \atheva & \lehamoc & GRU \nn\\
        \hline \hline
        Single SSC model & 1-4~min & 2-3~s & 3~ms \\
        {\tt emcee}\tnote{*} & -- & 20~d & 20~min\\
        {\tt UltraNest} & -- & -- & 21~hr \\
        \hline
    \end{tabular}
    \begin{tablenotes} 
    \item[*] For 48 walkers and 10,000 steps for each chain. 
    \end{tablenotes}
    \end{threeparttable}
\end{table}

\section{Conclusions}\label{sec:conclusions}

We have presented a proof-of-concept study of blazar SED modeling that relies on the use of neural networks (\nns) and Bayesian inference. We have demonstrated how computationally expensive numerical models can be substituted with trained \nns. By testing 3 typical configurations of neurons we conclude that a GRU \nn offers optimal results for the astrophysical problem at hand. We have tested the efficiency of our approach against simulated datasets and observational data.
Our results demonstrate the big leaps in computing efficiency that can be achieved compared to state-of-the-art radiative numerical codes. This in turn makes the process of fitting  about 1,000 times faster, enabling the use of not only typical MCMC methods but also nesting sampling algorithms like in {\tt UltraNest}.These findings will provide extra motivation in the development and testing of different \nns or neurons and experimenting with more complex configurations.

A natural next step would be the training of \nns against more complex radiative models that include lepto-hadronic processes. A typical example is the case of \hsp \,  where lepto-hadronic models computed with \atheva have only been tested by eye against the data \citep{2020ApJ...899..113P} -- see also \cite{Petro2015} for similar applications. Given the large number of parameters (at least 11) and the execution time (tens of minutes, even with faster codes like \lehamoc) a statistical fitting with Bayesian methods is prohibitive. In lepto-hadronic models the number of PDEs that need to be solved increases from two to at least five. Moreover, the PDE describing the evolution of each particle species in the source is non-linearly coupled to the PDEs of other species. Due to the non-linear coupling small change in the model parameters may lead to drastic changes in the resulting SED. Complexity of lepto-hadronic models is further increased by the fact that not all equilibrium solutions of the problem are constant in time. There are regions of the high-dimension parameter space, as extensively discussed by \cite{2020MNRAS.495.2458M}, that produce oscillatory equilibrium solutions \citep[][]{PM12, Mastichiadis_2005, 2018MNRAS.477.2917P}, known as limit cycles in non-linear dynamics \citep{strogatz:2000}. The methodology presented here can also be extended to other astrophysical problems, like spectral modeling of X-ray pulsars \citep[see examples of available numerical models][]{2016ApJ...831..194W,2017ApJ...835..129W,2022ApJ...939...67B} or GRBs \citep[e.g.][]{2023ApJ...950...28R}.\\

\textit{Code Availability. --} The results of our work will be made available in a GitHub repository (\url{https://github.com/tzavellas/blazar_ml}) upon acceptance of the paper, this includes; (a) the \nn and accompanied code produced to train them, (b) code for visualization of results in python and jupyter notebooks with instructions, and (c) part of the \atheva datasets that can be used for evaluation and plotting examples. \\

\textit{Note. --}  While we were finalizing this work, the paper by \citet{2023arXiv231102979B} appeared. While the main concept and the radiation model are the same, the two studies are independent and complementary. They focus on a different \nn, evaluation and modeling examples. 

\begin{acknowledgements}
M.P. acknowledge support from the Hellenic Foundation for Research and Innovation (H.F.R.I.) under the ``2nd call for H.F.R.I. Research Projects to support Faculty members and Researchers'' through the project UNTRAPHOB (Project ID 3013). GV acknowledges support by H.F.R.I. through the project ASTRAPE (Project ID 7802). 
\end{acknowledgements}

%
%

\bibliographystyle{aa} 
\bibliography{ref.bib}

\appendix

\section{Corner plots} 

\begin{figure}
\centering
\includegraphics[width=0.99\textwidth]{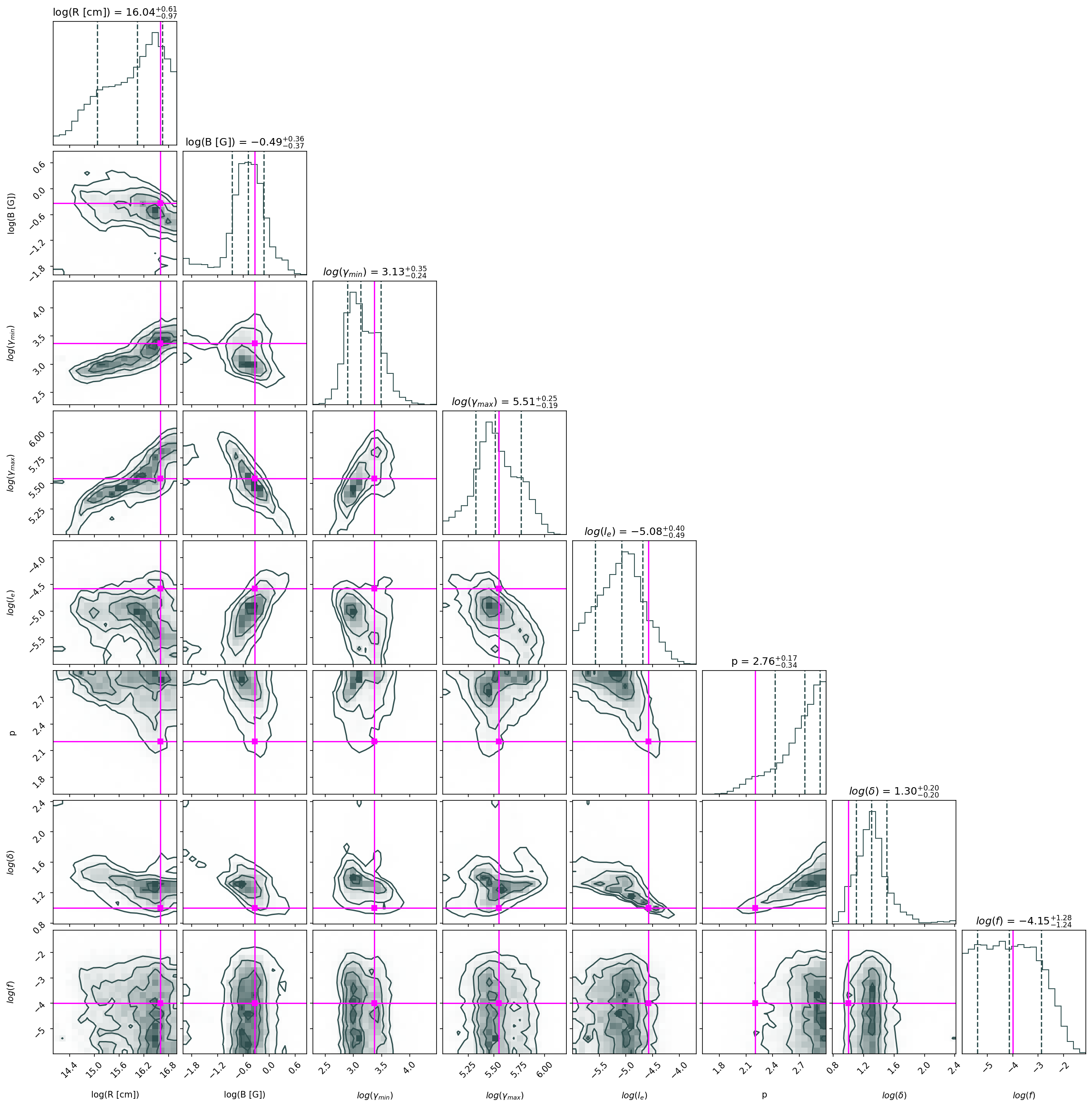} 
\caption{Corner plot showing the posterior distributions of the parameters of the GRU \nn model presented in Fig. \ref{fig:fit_SED} (left). Contours denote Parameters of the initial model are marked with magenta lines.}
\label{fig:corner-fake12}
\end{figure}

\begin{figure*}
\centering
\includegraphics[width=0.99\textwidth]{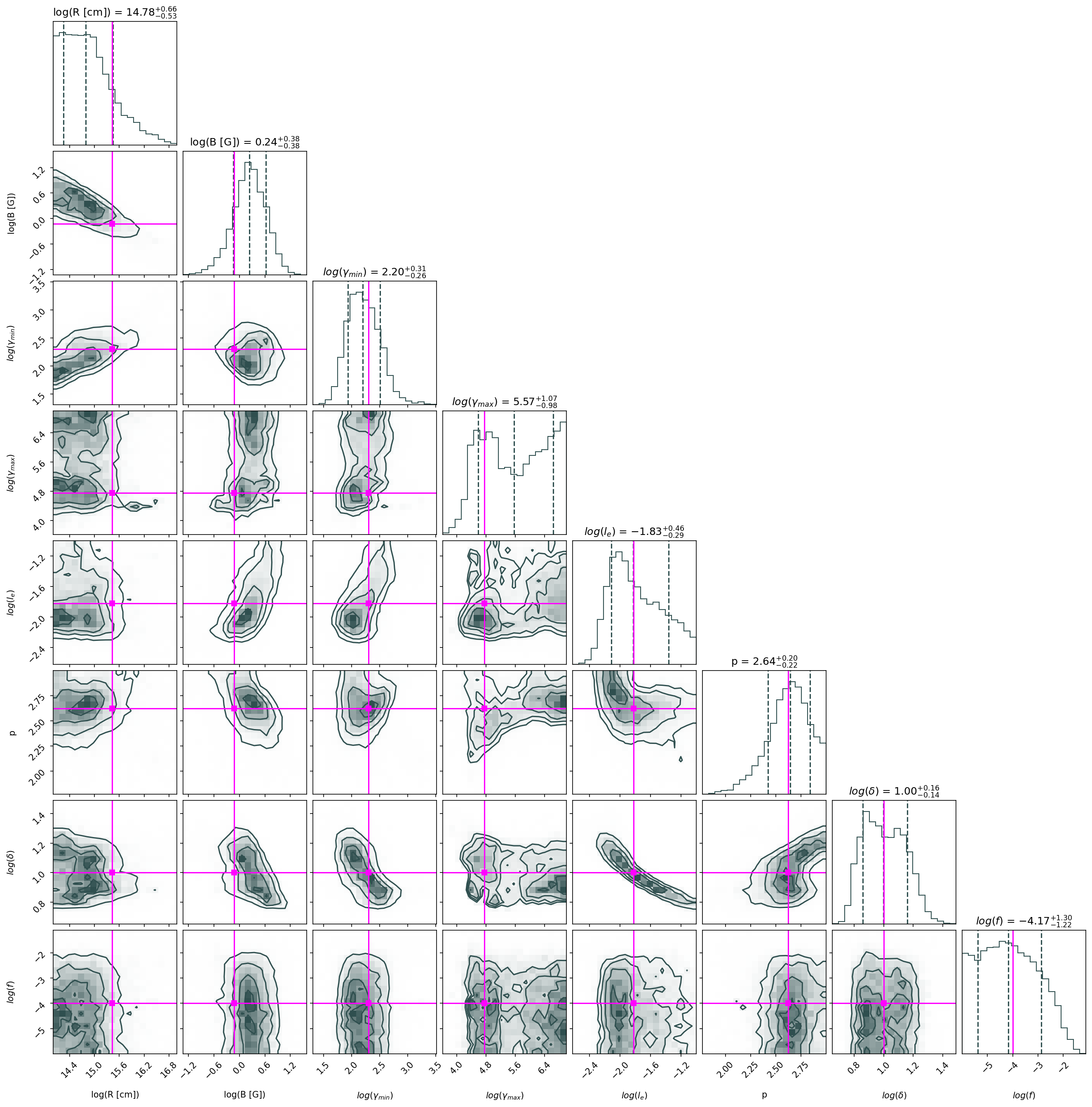} 
\caption{Corner plot showing the posterior distributions of the parameters of the GRU \nn model presented in Fig. \ref{fig:fit_SED} (right). Parameters of the initial model are marked with magenta lines.}
\label{fig:corner-fake27}
\end{figure*} 

\begin{figure*}
\centering
\includegraphics[width=0.99\textwidth]{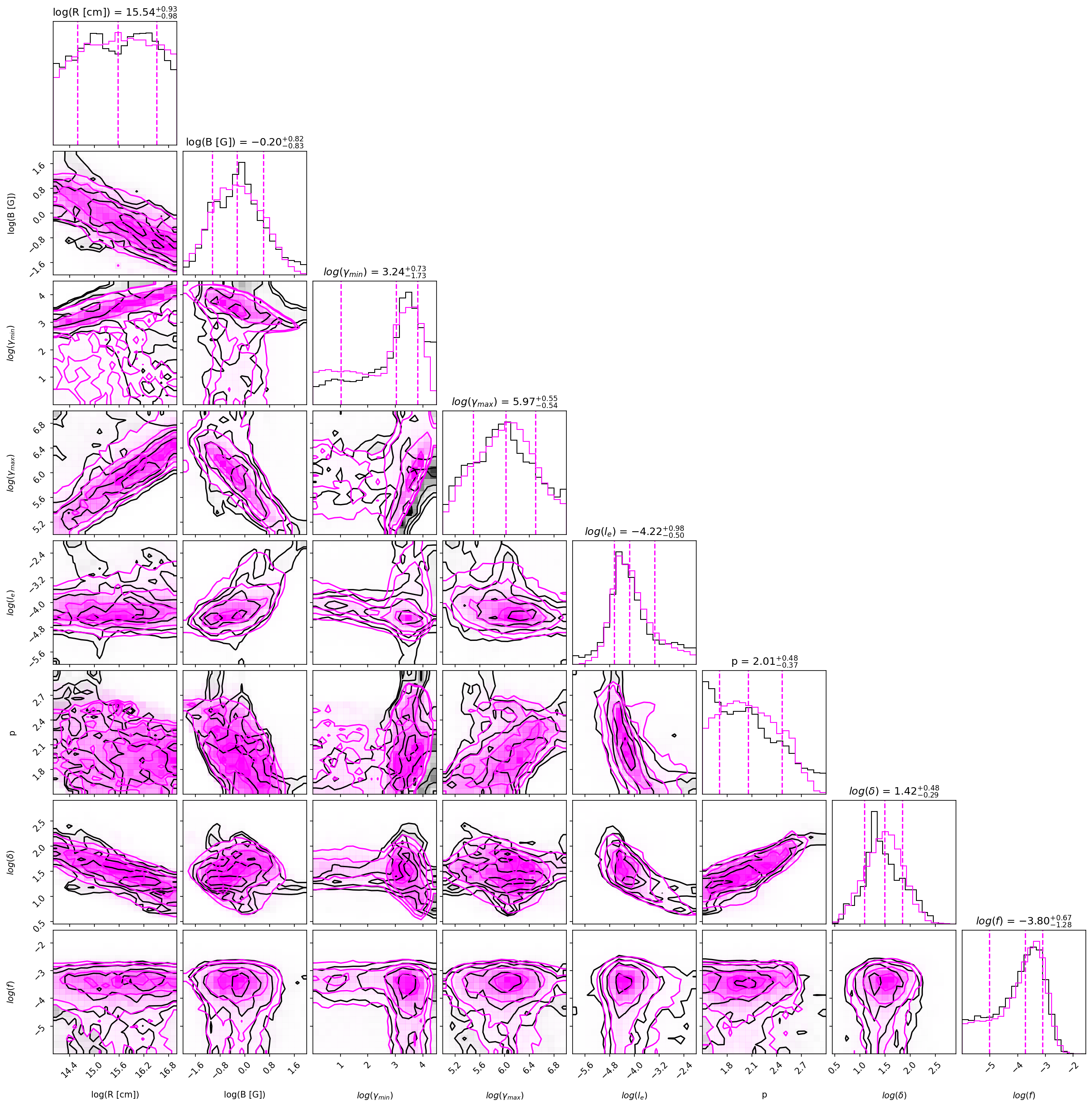} 
\caption{Corner plot showing the posterior distributions of the GRU \nn model parameters for \hsp as derived from {\tt emcee}. Dashed lines in the histograms indicate the median and 68 \% range of values. With magenta color we overplot the corner plot of the SSC model fitted by \lehamoc as presented by \citet{2023arXiv230806174S}.}
\label{fig:corner-ssc}
\end{figure*}

\begin{figure*}
\centering
\includegraphics[width=0.99\textwidth]{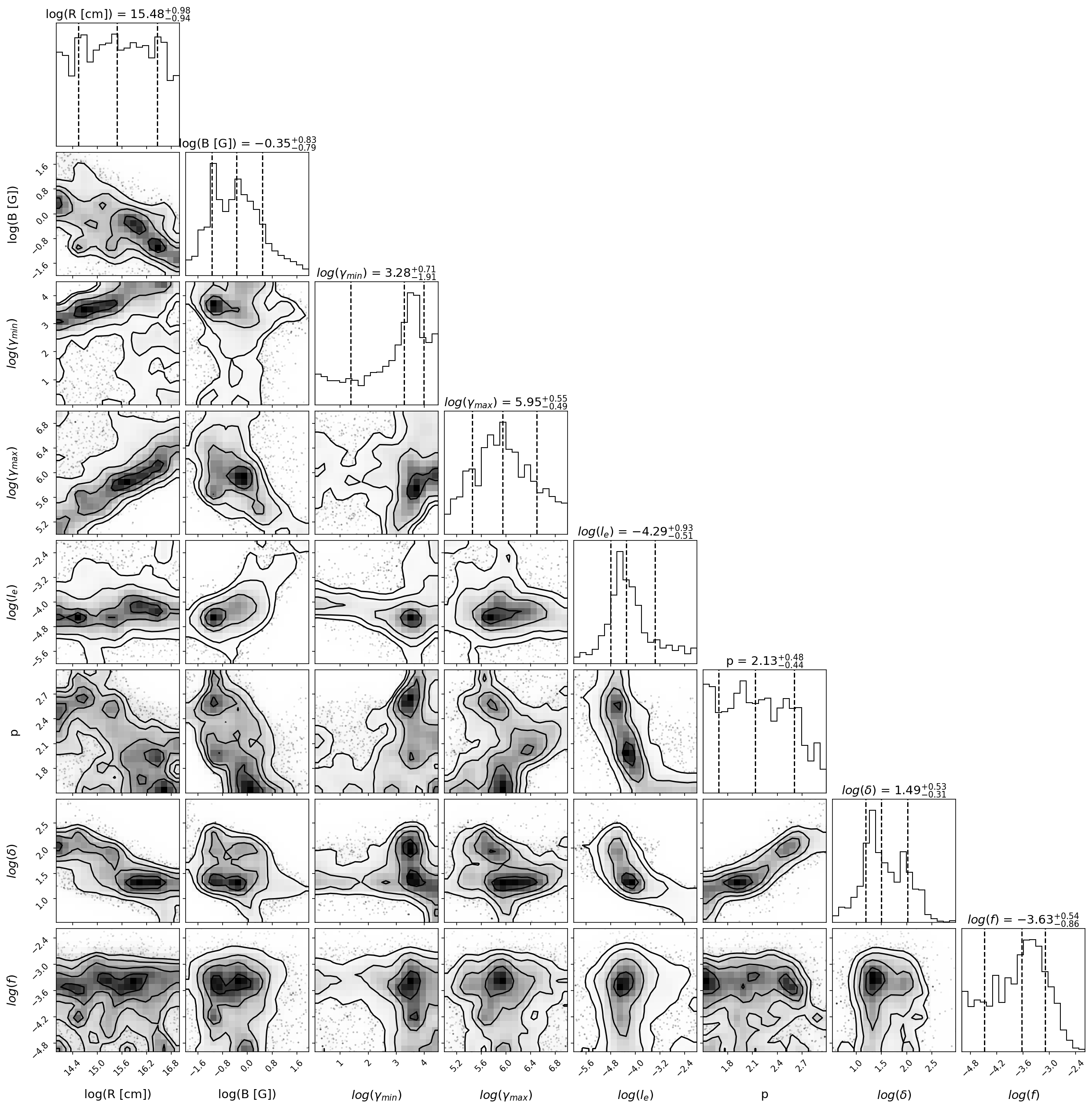} 
\caption{Corner plot showing the posterior distributions of the GRU \nn model parameters for \hsp as derived from {\tt UltraNest}. Dashed lines in the histograms indicate the median and 68 \% range of values.}
\label{fig:corner-ssc-ultra}
\end{figure*}

\end{document}